\documentclass[12pt]{iopart}
\usepackage{iopams}
\newcommand{\Eins}
           {\;\smash{\raisebox{-0.5ex}{$\!\!\stackrel{\!\mbox{1}
            \hspace{-0.4ex}\rule[0.0ex]{0.06ex}{1.60ex}}{ }$}}}
\newcommand{\Rangle}{\rangle\hspace{-0.4ex}\rangle}

\newtheorem{prop}{Proposition}
\newtheorem{theorem}{Theorem}

\begin{document}

\title[Linear energy bounds]{Linear energy bounds for Heisenberg
spin systems}

\author{Heinz-J\"urgen Schmidt
}

\address{\dag\ Universit\"at Osnabr\"uck, Fachbereich Physik,
Barbarastr. 7, 49069 Osnabr\"uck, Germany}

\begin{abstract}
Recently obtained results on linear energy bounds are generalized
to arbitrary spin quantum numbers and coupling schemes. Thereby
the class of so--called independent magnon states, for which the
relative ground--state property can be rigorously established,
is considerably enlarged. We still require that the
matrix of exchange parameters has constant row sums, but this can be
achieved by means of a suitable gauge and need not be considered as a
physical restriction.
\end{abstract}


\maketitle

\section{Introduction}

For ferromagnetic spin systems the ground--state
$\left|\uparrow \uparrow \ldots \uparrow  \rangle\right.$
and the first few excited states, called magnon states, are
well--known and extensively investigated, see e.~g.~\cite{Mat}.
For anti--ferromagnetic (AF) coupling the state
$\left|\uparrow \uparrow \ldots \uparrow  \rangle\right.$
will be the state of highest energy, and could be called
the "anti--ground--state". Also the magnon states which have a large
total spin quantum number $S$ are still eigenstates of
the Heisenberg Hamiltonian, but seem to be of less physical importance
at first glance, since in thermal equilibrium they are dominated by the
low--lying eigenstates. However, these states will become ground--states
if a sufficient strong magnetic field \textsf{H} is applied, since
the Zeeman term in the Hamiltonian will give rise to a maximal energy shift
of $-  \mu_B g S  \textsf{H} $. Hence the magnon states are yet important
for the magnetization curve ${\cal M}(\textsf{H})$, especially at low temperatures.\\
What can then be said about the energies of the magnon states in general?
The anti--ground state (or "magnon vacuum") has the energy
$E_0 = J s^2 $ and the total magnetic quantum number $M=Ns$.
Here $N$ denotes the number of spins with individual spin quantum number $s$
and $J$ denotes the sum of all exchange parameters of the spin system, see
section 2 for details. The $1$--magnon states ly in the subspace with
$M=Ns-1$. Their energies can be calculated, up to a constant shift and a factor $2s$,
as the eigenvalues of the symmetric $N\times N$--matrix of exchange parameters,
which can be done exactly in most cases.
Let $E_1^{\min{}}$ denote the smallest of
these energies. More generally, we will write
$E_a^{\min{}}$ for the minimal energy
within the subspace with total magnetic quantum number $M=Ns-a$.
It turns out that typically the graph of
$a\mapsto E_a^{\min{}}$ will be an approximate
parabola with positive curvature, see \cite{JM}\cite{SSL1}.
However, there are exceptions
to this rule, see \cite{Wal}\cite{SSL2},
one exception being given by the recently discovered
"independent magnon states" \cite{SSRS}\cite{SHSRS}. Here, for some small values of $a$, we have
\begin{equation}\label{1}
E_a^{\min{}} = (1-a) E_0 + a E_1^{\min{}},
\end{equation}
i.~e.~$a\mapsto E_a^{\min{}}$ is locally an affine function.
The existence of states satisfying (\ref{1}) is not just a curiosity
but has interesting physical consequences with respect to magnetization:
Since the Zeeman term is also linear in $a$, the independent magnon states
simultaneously become ground--states at the saturation value of the
applied magnetic field. One would hence observe a marked jump
in the magnetization curve ${\cal M}(\textsf{H})$ at zero temperature, see
the discussion in \cite{SSRS}. Other examples of spin systems exhibiting
jumps in the magnetization curve due to linear parts of the energy spectrum
are known, see
\cite{Mil}\cite{MHSKU}\cite{HMT}\cite{KOK}\cite{SR} and \cite{CB}.\\
The independent magnon states which satisfy (\ref{1}) can be analytically
calculated. In order to rigorously prove that their energy eigenvalues
are minimal within the subspaces with $M=Ns-a$, one could try to prove
a general inequality of the form
\begin{equation}\label{2}
E_a^{\min{}} \ge (1-a) E_0 + a E_1^{\min{}}
\end{equation}
for all $a=0,\ldots,2Ns$ and AF-coupling. Geometrically, (\ref{2})
means that energies in the plot of $E$ versus $a$ ly on or above
the line joining the first two points with $E_0$ and $E_1$.\\
A proof of (\ref{2}) is given in \cite{SSRS} only for the special case
of $s=\frac{1}{2}$ and certain homogeneous coupling schemes.
This is an unsatisfying situation since the independent magnon states
constructed in \cite{SSRS}\cite{SHSRS} can be defined for any $s$ and their
minimal energy property is numerically established without any doubt.\\
Thus this article is devoted to the generalization of the quoted proof
to arbitrary $s$ and coupling schemes. The only assumption we need is that
the exchange parameters $J_{\mu\nu}$ have equal signs. For AF--coupling,
i.~e.~$J_{\mu\nu}\ge 0$, we obtain (\ref{2}). The ferromagnetic case
$J_{\mu\nu}\le 0$ is completely analogous and yields
\begin{equation}\label{3}
E_a^{\max{}} \le (1-a) E_0 + a E_1^{\max{}},
\end{equation}
with self-explaining notation. Hence it will not be necessary to
consider the ferromagnetic case separately in the rest of this article.\\
As remarked before, it is an obvious benefit of the generalisation of the
quoted proof to rigorously establish the minimal energy property
of the independent magnon states for arbitrary $s$. Moreover, it is
now easy to extend the construction of independent magnon states to
coupling schemes with different exchange parameters. For example,
one could consider a cuboctahedron with two different exchange constants:
$J_1>0$ for the bonds within two opposing squares and $J_2$ satisfying
$0<J_2<J_1$ for the remaining bonds. This coupling scheme would admit the
same independent magnon states as those considered in \cite{SSRS}.\\
The technique of the proof of the generalized inequality is essentially
the same as that of the old one. The generalization to arbitrary $s$ is achieved
by replacing every spin $s$ by a group of $2s$ spins $\frac{1}{2}$ and the coupling
between two spins by a uniform coupling between the corresponding groups. The
energy eigenvalues of the new system include the eigenvalues of the old one.
In this way the proof can be reduced to the case of $s=\frac{1}{2}$.\\
In the next step we embed the Hilbert space of the spin $\frac{1}{2}$
system into some sort of bosonic Fock space for magnons and compare the
Heisenberg Hamiltonian with that for the ideal magnon gas. The difference
of these two Hamiltonians has two components of different origin: First,
there occurs some (positive definite) term due to a kind of "repulsion"
between magnons. Only here the AF-coupling assumption is needed. Second, the
"kinetic energy" part (or XY--part) of the magnon gas Hamiltonian produces some
unphysical states, due to the fact that the magnon picture is only an
approximation of the real situation. The necessary projection onto the
physical states further increases the ground state energy. Both components
introduce a $>$--sign in (\ref{2}). As far as the Heisenberg spin system
can be {\it exactly} viewed as an ideal magnon gas, we have an $=$--sign
in (\ref{2}) as for independent magnons. The analogy of a antiferromagnet
in a strong magnetic field with a repulsive Bose gas is well-known, see
for example \cite{BB}\cite{Glu}\cite{SSS}.\\
The paper is organized as follows: Section 2 contains the pertinent notation
and definitions, section 3 the main theorem together with its proof in two
steps (sections 3.1 and 3.2) and section 4 a short discussion.

\section{Notation and Definitions}

We consider systems with $N$ spin sites, individual spin quantum number $s$
and Heisenberg Hamiltonian
\begin{equation}\label{2.1}
H = \sum_{\mu,\nu=1}^N J_{\mu\nu}\;{\bf s}_\mu\cdot{\bf s}_\nu.
\end{equation}
Here
${\bf s}_\mu =\left({\bf s}_\mu^{(1)},{\bf s}_\mu^{(2)},{\bf s}_\mu^{(3)}\right)$
is the (vector) spin operator at site $\mu$ and $J_{\mu\nu}$ is the exchange parameter
determining the strength of the coupling between sites $\mu$ and $\nu$.
$J_{\mu\nu}$ will be considered as the entries of a real
$N\times N$-matrix $\mathbf J$.
As usual,
\begin{equation}\label{2.2a}
{\bf S}^{(i)}\equiv \sum_\mu   {\bf s}_\mu^{(i)} \quad (i=1,2,3)
\end{equation}
and
\begin{equation}\label{2.2b}
 {\bf s}_\mu^{\pm} \equiv  {\bf s}_\mu^{(1)}\pm i{\bf s}_\mu^{(2)},
 \quad \mu=1,\ldots,N.
\end{equation}
The Hilbert space which is the domain of definition of the various operators
considered will be denoted by ${\cal H}(N,s)$. It can be identified with
the $N$--fold tensor product
\begin{equation}\label{2.2c}
{\cal H}(N,s)  = \bigotimes_{i=1}^{N}{{\cal H}(1,s)}.
\end{equation}
Note that the exchange parameters $J_{\mu\nu}$ are not uniquely determined
by the Hamiltonian $H$ via (\ref{2.1}). Different choices of the $J_{\mu\nu}$
leading to the same $H$ will be referred to as different ``gauges".\\
First, the anti--symmetric part of $\mathbf J$ does not enter into (\ref{2.1})
and could be chosen arbitrarily.
However, throughout this article we will choose $J_{\mu\nu}=J_{\nu\mu}$,
i.~e.~consider $\mathbf J$ as a symmetric matrix.
Second, the diagonal part of $\mathbf J$ is not fixed by (\ref{2.1}).
Since  ${\bf s}_\mu\cdot{\bf s}_\mu =s(s+1)\Eins$ we may choose arbitrary
diagonal elements  $J_{\mu\mu}$ without changing $H$, as long as their sum
vanishes, $\mbox{Tr}{\mathbf J}=0$. The usual gauge chosen throughout the
literature is  $J_{\mu\mu}=0,\; \mu=1,\ldots,N,$ which will be called the
``zero gauge". In this article, however, we will choose another gauge,
called ``homogeneous gauge", which is defined by the condition that the row sums
\begin{equation}\label{2.3}
J_\mu \equiv \sum_\nu  J_{\mu\nu}
\end{equation}
will be independent of $\mu$.
Of course, there exist spin systems which admit both gauges simultaneously,
e.~g.~homogeneous spin rings. These systems will be called ``weakly homogeneous".
We will see that the condition of weak homogeneity used in previous articles
\cite{SSL1} \cite{SSRS} is largely superfluous and can be replaced by the homogeneous
gauge (but see section 4).\\
Note that the eigenvalues of $\mathbf J$
may non--trivially depend on the gauge. The homogeneous gauge has the
advantage that energy eigenvalues in the 1--magnon--sector are simple
functions of the eigenvalues of $\mathbf J$, see below.
The quantity
\begin{equation}\label{2.4}
J\equiv \sum_{\mu\nu}  J_{\mu\nu}
\end{equation}
is gauge--independent. If exchange parameters satisfying
$\tilde{J}_{\mu\nu}=\tilde{J}_{\nu\mu}$
are given in the zero gauge, the corresponding
parameters $J_{\mu\nu}$ in the homogeneous gauge are obtained as follows:
\begin{equation}\label{2.5a}
J_{\mu\nu} \equiv \tilde{J}_{\mu\nu} \mbox{ for } \mu \neq \nu,
\end{equation}
\begin{equation}\label{2.5b}
J_{\mu\mu} \equiv \textstyle\frac{1}{N}\; J - \tilde{J}_\mu.
\end{equation}
It follows that
\begin{equation}\label{2.6}
j\equiv J_\mu =\sum_\nu \tilde{J}_{\mu\nu} + J_{\mu\mu} = \textstyle\frac{J}{N}.
\end{equation}
Since $H$ commutes with ${\bf S}^{(3)}$, the eigenspaces
${\cal H}_a$ of ${\bf S}^{(3)}$ with eigenvalues
$M=Ns-a,\; a=0,1,\ldots,2Ns,$ are invariant under the action of $H$.
${\cal H}_a$ will be called the $a$--magnon--sector. Let
${\cal P}_a$ denote the projection onto  ${\cal H}_a$ and
\begin{equation}\label{2.7}
H_a\equiv {\cal P}_a H {\cal P}_a.
\end{equation}
An orthonormal basis of ${\cal H}_a$ is given by the product states
$\| m_1,m_2,\ldots,m_N \Rangle$ satisfying
\begin{equation}\label{2.8}
{\bf s}_\mu^{(3)}\| m_1,m_2,\ldots,m_N \Rangle = m_\mu
\| m_1,m_2,\ldots,m_N \Rangle,
\end{equation}
where $m_\mu$ can assume the  $2s$ values
\begin{equation}\label{2.9}
m_\mu = s, s-1,\ldots, -s.
\end{equation}
In the case of $s=\frac{1}{2}$,
$m_\mu\in\left\{\frac{1}{2},-\frac{1}{2}\right\}   \equiv
\left\{\uparrow,\downarrow\right\}$
and the state (\ref{2.8}) can be uniquely specified by the
ordered set $|n_1,\ldots,n_a\rangle$ of  $a$ spin sites
$\mu$ with $m_\mu =-\frac{1}{2}$. We will use both notations
equivalently:
\begin{equation}\label{2.10}
|n_1,\ldots,n_a\rangle \equiv \| m_1,m_2,\ldots,m_N \Rangle.
\end{equation}
For example,
\begin{equation}\label{2.11}
| 1,3,4\rangle = \| \downarrow \uparrow \downarrow \downarrow \uparrow  \Rangle,
\; a=3,\; N=5.
\end{equation}
We now consider again arbitrary $s$.
The subspace ${\cal H}_1$
is $N$--dimensional and, similarly as above, its basis vectors
$\| m_1,m_2,\ldots,m_N \Rangle$ may be denoted
by $| n\rangle,\; n=1,\ldots,N,$ if $n$ denotes the site with lowered spin,
i.~e.~$m_\mu = s - \delta_{\mu n},\; \mu=1,\ldots,N.$ \\
Consider
\begin{eqnarray}\label{2.12}
H_1 & = & {\cal P}_1 \left(
\sum_{\mu\nu} J_{\mu\nu} {\bf s}_\mu^{(3)}{\bf s}_\nu^{(3)}  +
\textstyle\frac{1}{2}
\sum_{\mu\nu} J_{\mu\nu} ({\bf s}_\mu^{+}{\bf s}_\nu^{-} +
{\bf s}_\mu^{-}{\bf s}_\nu^{+})
\right)   {\cal P}_1 \\
 &\equiv& H_1^Z + H_1^{XY},
\end{eqnarray}
and
\begin{eqnarray}\label{2.13}
H_1^Z |n\rangle & = &
\left(
\sum_{\mu\nu} J_{\mu\nu}  (s-\delta_{\mu n})(s-\delta_{\nu n})
\right) |n\rangle\\
 & = &
 \left(
 s^2 Nj -2sj +J_{nn}
\right) |n\rangle.
\end{eqnarray}
Similarly, we obtain after some calculation
\begin{equation}\label{2.14}
H_1^{XY}|n\rangle
=
2s \sum_{m,m\neq n} J_{n m} |m\rangle
+
(2s-1) J_{nn}|n\rangle.
\end{equation}
Hence
\begin{equation}\label{2.15}
H_1 = (s^2 Nj -2sj)\Eins_{{\cal H}_a} + 2s {\mathbf J}
\end{equation}
and the eigenvalues of $H_1$ are
\begin{equation}\label{2.16}
E_\alpha = s^2 Nj + 2s (j_\alpha -j),
\end{equation}
if $j_\alpha,\; \alpha =1,\ldots,N,  $ are the eigenvalues
of ${\mathbf J}$.
This simple relation between $H_1$ and ${\mathbf J}$
only holds in the homogeneous gauge.
Note further that, due to the homogeneous gauge,
$j$ is one of the eigenvalues of
${\mathbf J}$, the corresponding eigenvector
having constant entries.\\
We denote by $j_{\min{}}$ the minimal eigenvalue
of ${\mathbf J}$ and by $E_1^{\min{}}$ the corresponding minimal eigenvalue
of $H_1$.

\section{The Main Result}

\begin{theorem}
Consider a spin system with AF-Heisenberg coupling scheme
and homogeneous gauge, i.~e.~
\begin{equation}\label{3.17}
j\equiv \sum_\nu J_{\mu\nu}
\end{equation}
being independent of $\mu$ and
\begin{equation}\label{3.18}
J_{\mu\nu}\ge 0 \mbox{ for } \mu\neq\nu.
\end{equation}
Then the following operator inequality holds:
\begin{equation}\label{3.19}
H_a \ge
\left(
Njs^2 -2sa(j-j_{\min{}})
\right)
\Eins_{{\cal H}_a}
\end{equation}
for all $a=0,1,\ldots,2Ns$.
\end{theorem}
The rest of this section is devoted to the proof of this theorem.

\subsection{Reduction to the case $s=\frac{1}{2}$}

We will construct another Hamiltonian
\begin{equation}\label{3.20}
\widehat{H}=\sum_{\alpha,\beta=1}^{\widehat{N}}
\widehat{J}_{\alpha\beta}\;\widehat{\bf s}_\alpha\cdot\widehat{\bf s}_\beta
\end{equation}
acting on the Hilbert space
$\widehat{\cal H}={\cal H}(2Ns,\frac{1}{2})$,
i.~e.~$\widehat{N}=2Ns$ and $\widehat{s}=\frac{1}{2}$.
Intuitively, every spin site with spin $s$ is replaced by a group
of $2s$ spin sites with spin $\frac{1}{2}$ and
the coupling between spin sites is extended to a uniform
coupling between groups, see figure 1.\\
\setlength{\unitlength}{1cm}
\begin{figure}[h]
\begin{picture}(15,6)
\put(1,2.5){\circle*{0.4}}
\put(0.8,2){\mbox{$\mu$}}
\put(5,2.5){\circle*{0.4}}
\put(4.8,2){\mbox{$\nu$}}
\put(1,2.5){\line(1,0){4}}
\put(6.5,2.4){\mbox{$\Longrightarrow$}}
\multiput(8,1)(0,1){4}{\circle*{0.2}}
\put(7.8,0.5){\mbox{$\alpha$}}
\multiput(12,1)(0,1){4}{\circle*{0.2}}
\put(11.8,0.5){\mbox{$\beta$}}
\multiput(8,1)(0,1){4}{\line(1,0){4}}
\multiput(8,1)(0,1){3}{\line(4,1){4}}
\multiput(8,4)(0,-1){3}{\line(4,-1){4}}
\multiput(8,1)(0,1){2}{\line(2,1){4}}
\multiput(8,4)(0,-1){2}{\line(2,-1){4}}
\put(8,1){\line(4,3){4}}
\put(8,4){\line(4,-3){4}}
\end{picture}
\caption[Reduction]{Reduction to the case $s=\frac{1}{2}$
by replacing single spins by groups of $2s$ spins $\frac{1}{2}$.}
\end{figure}
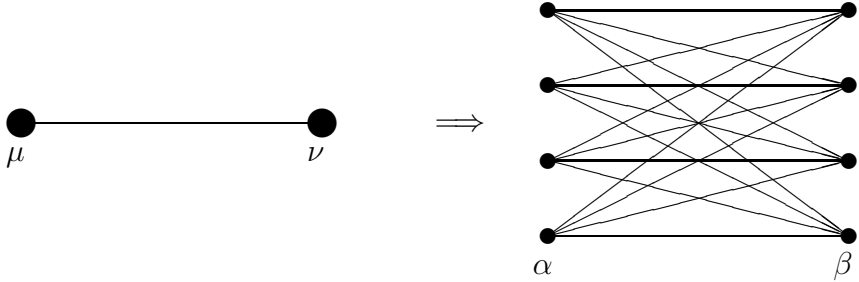

Formally, we set
\begin{equation}\label{3.21a}
\alpha\equiv (\mu,i),\;i=1,\ldots,2s
\end{equation}
and
\begin{equation}\label{3.21b}
\widehat{J}_{\alpha\beta}=\widehat{J}_{(\mu,i)(\nu,j)}
\equiv
J_{\mu\nu},\;i,j=1,\ldots,2s.
\end{equation}
The new matrix $\widehat{\mathbf{J}}$
satisfies the homogeneous gauge condition if   $\mathbf{J}$ does.\\
According to the well--known theory of the coupling of angular momenta
or spins the tensor product spaces
\begin{equation}\label{3.22}
{\cal H}(2s, \textstyle\frac{1}{2})    =
\bigotimes_{i=1}^{2s}{\cal H}(1,\textstyle\frac{1}{2})
\end{equation}
can be decomposed into eigenspaces of
$\left(
\sum_{i=1}^{2s} {\bf s}_i
\right)^2
$
with eigenvalues\\
$S(S+1),\; S=s,s-1,\ldots,
\left\{
\begin{array}{r@{\quad \mbox{if}\quad}l}
0 & 2s \mbox{ even }\\
\frac{1}{2} & 2s \mbox{ odd }
\end{array}
\right. $
The eigenspace with the maximal $S=s$
will be denoted by
${\cal K}_s$ and the projector onto this eigenspace by
${\mathbf P}_s$. ${\cal K}_s$ is isomorphic to ${\cal H}(1,s)$.
This isomorphism can be chosen such that the following
isometric embedding
\begin{equation}\label{3.23}
{\sf j}_s : {\cal H}(1,s) \longrightarrow {\cal K}_s
\hookrightarrow
{\cal H}(2s,\textstyle\frac{1}{2})
\end{equation}
satisfies
\begin{equation}\label{3.24}
{\sf j}_s {\bf s}{\sf j}_s^\ast = {\mathbf P}_s
\left(
\sum_{i=1}^{2s}{\bf s}_i
\right) {\mathbf P}_s.
\end{equation}
Let ${\sf j}$ denote the tensor product of the ${\sf j}_s$
\begin{equation}\label{3.25}
{\sf j}: {\cal H}(N,s) \longrightarrow {\cal H}(2Ns,\textstyle\frac{1}{2})
\end{equation}
and
${\mathbf P}=\bigotimes_{\nu =1}^{N}{\mathbf P}_s$
the corresponding projector onto
$\bigotimes_{\nu =1}^{N} {\cal K}_s$
which commutes with $\widehat{H}$.
Then it follows from (\ref{3.24}) that
\begin{equation}\label{3.26}
{\sf j} H {\sf j}^\ast = {\mathbf P} \widehat{H}{\mathbf P}.
\end{equation}
In other words, $H$ may be viewed as the restriction of $\widehat{H}$
onto the subspace of states with maximal spin
$S=s$ within the groups. The eigenvalues of  $H$
form a subset of the eigenvalues of  $\widehat{H}$.\\

The relation (\ref{3.21b}) between the exchange parameters
may be written in matrix form as
\begin{equation}\label{3.27}
\widehat{\mathbf J}={\mathbf J} \otimes  {\mathbf E},
\end{equation}
where ${\mathbf E}$ is the $2s\times 2s$--matrix completely filled with $1$'s.
The eigenvalues of ${\mathbf E}$ are $2s$ and $0$, the latter being
$(2s-1)$-fold degenerate, hence the eigenvalues of the
$\widehat{\mathbf J}$--matrix satisfy
\begin{equation}\label{3.28}
\widehat{j}_\alpha = 2s j_\alpha \mbox{ or } 0.
\end{equation}
It will be illustrative to check the spectral inclusion property
for some known eigenvalues of Heisenberg Hamiltonians.\\
The eigenvalue of $H$ for the magnon vacuum state
$ | \uparrow \uparrow \ldots \uparrow \rangle $  is
\begin{equation}\label{3.29}
E_0= Njs^2.
\end{equation}
For $\widehat{H}$ we analogously have
\begin{equation}\label{3.30}
\widehat{E_0}= (2sN)\widehat{j} (\textstyle\frac{1}{2})^2,
\end{equation}
which is identical with (\ref{3.29})
by $\widehat{j}=2sj$.
Similarly, the eigenvalues of $H$ in the $1$--magnon--sector
$M=Ns-1$ are
\begin{equation}\label{3.31}
E_\alpha=jNs^2 +2s(j_\alpha-j),
\end{equation}
cf.~(\ref{2.16}), hence
\begin{equation}\label{3.32}
\widehat{E_\alpha}=\widehat{j}\cdot 2sN\cdot  \textstyle\frac{1}{4}
+2\textstyle\frac{1}{2}(\widehat{j_\alpha}-\widehat{j}) ,
\end{equation}
which is identical with (\ref{3.31})
because of (\ref{3.28}).\\
Analogously to the cases considered it is easy to see that the bounds
of the {\it rhs} of (\ref{3.19})  are the same for
$H$ and $\widehat{H}$:
Since $\widehat{j}$ and $\widehat{j}_{\min{}}$
cannot be zero, they must satisfy
\begin{equation}\label{3.33}
\widehat{j}= 2s j,\; \widehat{j}_{\min{}} = 2s {j}_{\min{}},
\end{equation}
according to (\ref{3.28}).\\
Further, the spectrum of $H_a$ is contained in the spectrum of
$\widehat{H_a}$. Thus it suffices to prove (\ref{3.19})
for $\widehat{H}$, i.~e.~$s=\frac{1}{2}$.

\subsection{Embedding into the magnon Fock space}
Throughout this section we set  $s=\frac{1}{2}$.
Recall that
\begin{equation}\label{3.34}
{\cal H}(N,\textstyle\frac{1}{2})=\bigoplus_{a=0}^{N}{\cal H}_a
\end{equation}
denotes the decomposition of the Hilbert space of the system into
eigenspaces of ${\bf S}^{(3)}$ with eigenvalues
$M=\frac{1}{2} N-a$. \\
Let
\begin{equation}\label{3.35}
{\cal B}_a({\cal H}_1)\subset\bigotimes_{i=1}^{a}{\cal H}_1
\end{equation}
denote the completely symmetric subspace of the $a$--fold tensor product
of $1$--magnon--spaces and
\begin{equation}\label{3.36}
{\cal B}_a(T):{\cal B}_a({\cal H}_1)\longrightarrow{\cal B}_a({\cal H}_1)
\end{equation}
be the corresponding restriction of
$T\otimes \Eins\otimes \ldots \otimes  \Eins +
\ldots + \Eins\otimes\ldots\otimes \Eins\otimes T $
if $T:{\cal H}_1 \longrightarrow {\cal H}_1$ is
any linear operator.\\
Recall that a basis of ${\cal H}_a$ is given by the states
\begin{equation}\label{3.37}
|n_1,n_2,\ldots,n_a\rangle,\; 1\le n_1<n_2<\ldots<n_a\le N
\end{equation}
where the $n_i$ denote the lowered spin sites. Let
\begin{equation}\label{}
{\cal S}_a : \bigotimes_{i=1}^{a}{\cal H}_1 \longrightarrow {\cal B}_a({\cal H}_1)
\end{equation}
denote the ``symmetrizator", i.~e.~the sum over all permuted states
divided by the square root of its number. The assignement
\begin{equation}\label{3.39}
|n_1,n_2,\ldots,n_a\rangle \mapsto {\cal S}_a
|n_1\rangle \otimes |n_2\rangle \otimes \ldots \otimes   |n_a\rangle
\end{equation}
can be extended to an isometric embedding denoted by
\begin{equation}\label{3.40}
{\cal J}_a : {\cal H}_a \longrightarrow {\cal B}_a ({\cal H}_1).
\end{equation}
It satisfies ${\cal J}_a^\ast {\cal J}_a = \Eins_{{\cal H}_a}$.
Hence ${\sf P}_a\equiv {\cal J}_a^\ast {\cal J}_a$
will be a projector onto a subspace of
$ {\cal B}_a ({\cal H}_1)$ denoted by ${\cal I}_a({\cal H}_1)$.
Obviously,
\begin{equation}\label{3.41}
{\cal J}_a^\ast={\cal J}_a^\ast {\sf P}_a .
\end{equation}
The states contained in  ${\cal I}_a({\cal H}_1)$
are called ``physical states" since they are in $1:1$--correspondence
with the states in  ${\cal H}_a$.
The orthogonal complement of   ${\cal I}_a({\cal H}_1)$
contains ``unphysical states" like $|n\rangle\otimes|n\rangle$.
More general, it is easy to see that any superposition of product states
is orthogonal to  ${\cal I}_a({\cal H}_1)$
iff all product states of the superposition contain at least
one factor twice or more.\\
We define
\begin{equation}\label{3.42}
\widetilde{H}_a  \equiv   {\cal J}_a^\ast  {\cal B}_a ({\cal H}_1) {\cal J}_a.
\end{equation}
Recall that ${\cal H}_1$ was defined as the Hamiltonian in the $1$--magnon sector.\\
The main part of the remaining proof will consist of comparing
$\widetilde{H}_a$ with $H_a$. To this end, $H$ will be split into
a ``$Z$--part" and an ``$XY$--part" according to
\begin{eqnarray}\label{3.43}
H & = &
\sum_{\mu\nu}J_{\mu\nu}{\bf s}_\mu^{(3)}{\bf s}_\nu^{(3)} +
\textstyle\frac{1}{2}
\sum_{\mu\nu}J_{\mu\nu}
({\bf s}_\mu^{+}{\bf s}_\nu^{-}+{\bf s}_\mu^{-}{\bf s}_\nu^{+})
\\
 & \equiv & H^Z+H^{XY},
\end{eqnarray}
and, analogously, $H_a= H_a^Z+H_a^{XY}$ and
$\widetilde{H}_a= \widetilde{H}_a^Z+\widetilde{H}_a^{XY}$.

\begin{prop}
\begin{equation}
H_a^{XY} =  \widetilde{H}_a^{XY}.
\end{equation}
\end{prop}

{\bf Proof: }Let $|n_1,n_2,\ldots,n_a\rangle$ be an arbitrary basis vector
of ${\cal H}_a$. It suffices to consider a Hamiltonian of the form
\begin{equation}\label{3.44b}
H^{XY}=  \frac{1}{2}
({\bf s}_\mu^{+}{\bf s}_\nu^{-}+{\bf s}_\mu^{-}{\bf s}_\nu^{+}).
\end{equation}
Morover we need only consider the case $\mu<\nu$ since for $\mu=\nu$
the basis vectors are eigenvectors  both of
$H_a^{XY}$ and $\widetilde{H}_a^{XY}$ with eigenvalues $ \frac{1}{2}$.
We have to distinguish between four cases:
\begin{enumerate}
\item
$\mu,\nu \notin  \left\{ n_1,n_2,\ldots,n_a \right\}$:
\begin{eqnarray}\label{3.45}
H_a^{XY}|n_1,n_2,\ldots,n_a\rangle
& = &
\frac{1}{2}
({\bf s}_\mu^{+}{\bf s}_\nu^{-}+{\bf s}_\mu^{-}{\bf s}_\nu^{+})
|n_1,n_2,\ldots,n_a\rangle\\
& = & 0\\
& = &
\widetilde{H}_a^{XY}  |n_1,n_2,\ldots,n_a\rangle,
\end{eqnarray}
since $|n_1,n_2,\ldots,n_a\rangle$ is annihilated by
${\bf s}_\mu^{+}$ and ${\bf s}_\nu^{+}$.
\item
$\mu\in  \left\{ n_1,n_2,\ldots,n_a \right\}$, but
$\nu\notin  \left\{ n_1,n_2,\ldots,n_a \right\}$: \\
\begin{eqnarray}\nonumber
H_a^{XY} |n_1,\ldots,\mu\ldots,n_a\rangle
& = &
\textstyle\frac{1}{2}
{\bf s}_\mu^{+}{\bf s}_\nu^{-}
|n_1,\ldots,\mu\ldots,n_a\rangle
\\
& = &
\textstyle\frac{1}{2}
\mbox{Sort}
|n_1,\ldots,\nu,\ldots,n_a\rangle.
\end{eqnarray}
\begin{eqnarray}\nonumber
\widetilde{H}_a^{XY}
|n_1,\ldots,\mu\ldots,n_a\rangle
=
{\cal J}_a^\ast {\cal B}_a(H_1^{XY}){\cal J}_a
|n_1,\ldots,\mu\ldots,n_a\rangle&&\\  \nonumber
=
{\cal J}_a^\ast {\cal B}_a(H_1^{XY}){\cal S}_a
|n_1\rangle\otimes\ldots\otimes|n_a\rangle\\       \nonumber
=
\textstyle\frac{1}{2}{\cal J}_a^\ast  {\cal S}_a
\left(\right.
{\bf s}_\mu^{+}{\bf s}_\nu^{-}
|n_1\rangle\otimes\ldots\otimes|n_a\rangle &&\\   \nonumber
+\ldots +
|n_1\rangle\otimes\ldots\otimes
{\bf s}_\mu^{+}{\bf s}_\nu^{-}
|\mu\rangle \otimes\ldots\otimes |n_a\rangle &&\\    \nonumber
+\ldots +
|n_1\rangle\otimes\ldots\otimes
{\bf s}_\mu^{+}{\bf s}_\nu^{-}
|n_a\rangle
\left.\right) &&\\   \nonumber
=
\textstyle\frac{1}{2}{\cal J}_a^\ast  {\cal S}_a
|n_1\rangle\otimes\ldots\otimes|\nu\rangle\otimes\ldots\otimes|n_a\rangle
&&\\
=
\textstyle\frac{1}{2}
\mbox{Sort}|n_1,\ldots,\nu,\ldots,n_a\rangle.
\end{eqnarray}
Hence
$H_a^{XY} |n_1,\ldots,\mu\ldots,n_a\rangle =
\widetilde{H}_a^{XY} |n_1,\ldots,\mu\ldots,n_a\rangle$.
\item
The case
$\nu\in  \left\{ n_1,n_2,\ldots,n_a \right\}$, but
$\mu\notin  \left\{ n_1,n_2,\ldots,n_a \right\}$
is completely analogous.
\item
$\mu,\nu\in  \left\{ n_1,n_2,\ldots,n_a \right\}$:\\
In this case $H_a^{XY} |n_1,\ldots,\mu,\ldots,\nu,\ldots,n_a\rangle = 0$,
since the state is annihilated by
${\bf s}_\mu^-$ as well as by ${\bf s}_\nu^-$. On the other side
$
\widetilde{H}_a^{XY} =\\
\textstyle\frac{1}{2}{\cal J}_a^\ast  {\cal S}_a
\left(
|n_1\rangle\otimes\ldots\otimes|\nu\rangle\otimes\ldots
\otimes|\nu\rangle\otimes\ldots \otimes|n_a\rangle
+ \\
|n_1\rangle\otimes\ldots\otimes|\mu\rangle\otimes\ldots
\otimes|\mu\rangle\otimes\ldots \otimes|n_a\rangle
\right)=0,
$
since ${\cal J}_a^\ast  ={\cal J}_a^\ast   {\sf P}_a$
and ${\cal S}_a(\ldots)$ is orthogonal to the subspace
${\cal I}_a({\cal H}_1),$ see the remark after (\ref{3.41}).
\end{enumerate}
\hspace*{\fill}\rule{3mm}{3mm}  \\

\begin{prop}
\begin{equation}
H_a^{Z} \ge  \widetilde{H}_a^Z+\frac{1-a}{4}Nj.
\end{equation}
\end{prop}

{\bf Proof:} It turns out that the $|n_1,\ldots,n_a\rangle$
are simultaneous eigenvectors for $H_a^{Z}$ and $ \widetilde{H}_a^Z$:
First consider $H_a^{Z}$ and rewrite $|n_1,\ldots,n_a\rangle$ in the form
$\|m_1,\ldots,m_N\Rangle$ satisfying
\begin{equation}\label{3.51}
{\bf S}_\mu^{(3)}  \|m_1,\ldots,m_N\Rangle = m_\mu \|m_1,\ldots,m_N\Rangle.
\end{equation}
It follows that
\begin{eqnarray}\label{3.52}
H_a^{Z} \|m_1,\ldots,m_N\Rangle
&=&
\sum_{\mu\nu}J_{\mu\nu} m_\mu m_\nu   \|m_1,\ldots,m_N\Rangle \\
&\equiv&  \epsilon \|m_1,\ldots,m_N\Rangle.
\end{eqnarray}
We set
\begin{equation}\label{3.53}
a_\mu \equiv \textstyle\frac{1}{2}-m_\mu  \in \{0,1\}
\end{equation}
and obtain
\begin{eqnarray}\label{3.54}
\epsilon
&=&
\sum_{\mu\nu}J_{\mu\nu} m_\mu m_\nu \\
&=&
\textstyle\frac{1}{4} \left( \sum_{\mu\nu}J_{\mu\nu} \right)
-
\sum_{\mu\nu}J_{\mu\nu} a_\nu
+
\sum_{\mu\neq\nu}J_{\mu\nu} a_\mu a_\nu
+
\sum_{\mu}J_{\mu\mu} (a_\mu)^2\\
&=&
\textstyle\frac{1}{4} Nj -ja + \sum_{\mu}^\prime J_{\mu\mu}+\alpha,
\end{eqnarray}
where
\begin{equation}\label{3.55}
\alpha\equiv\sum_{\mu\neq\nu}  J_{\mu\nu} a_\mu a\nu  \ge 0,
\end{equation}
since $  J_{\mu\nu}\ge 0$ for $\mu\neq\nu$ by the assumption
of AF-coupling. $ \sum_{\mu}^\prime$ denotes the summation
over all $\mu$ with $a_{\mu}=1$. \\
Now consider
\begin{eqnarray}\label{^3.56}
\widetilde{H}_a^Z \| m_1,\ldots,m_N\Rangle
&=&
\widetilde{H}_a^Z |n_1,\ldots,n_a\rangle   \\   \nonumber
&=&
{\cal J}_a^\ast {\cal S}_a
\left(
H_1^Z|n_1\rangle \otimes |n_2\rangle \otimes \ldots \otimes|n_a\rangle
+\ldots \right.
\\
& &
\left.
+
|n_1\rangle \otimes |n_2\rangle \otimes \ldots \otimes H_1^Z|n_a\rangle
\right).
\end{eqnarray}
The terms $H_1^Z |n_i\rangle$ are special cases of
(\ref{3.52}),(\ref{3.54})
for $a=1$, hence
\begin{equation}\label{3.57}
H_1^Z |n_i\rangle = \textstyle\frac{1}{4}Nj -j + J_{n_i,n_i},
\end{equation}
since $\alpha=0$ in this case. We conclude
\begin{eqnarray}\label{3.58}
\widetilde{H}_a^Z \| m_1,\ldots,m_N\Rangle
& =&
\left(
a j (\textstyle\frac{N}{4}-1)+ \sum_\mu^\prime J_{\mu\mu}
\right)
\| m_1,\ldots,m_N\Rangle \\
&\equiv&
\delta  \| m_1,\ldots,m_N\Rangle.
\end{eqnarray}
Combining (\ref{3.52}),(\ref{3.54}) and (\ref{3.58}) yields
\begin{equation}\label{3.59}
\epsilon-\delta =\textstyle\frac{1-a}{4} Nj + \alpha \ge
\textstyle\frac{1-a}{4} Nj
\end{equation}
or
\begin{equation}\label{3.60}
H_a^Z - \widetilde{H}_a^Z \ge \textstyle\frac{1-a}{4} Nj \; \Eins.
\end{equation}
\hspace*{\fill}\rule{3mm}{3mm}  \\
The rest of the proof is straight forward.
Combining proposition 1 and 2 we obtain
\begin{equation}\label{3.61}
H_a \ge \widetilde{H}_a +
\textstyle\frac{1-a}{4}   Nj \; \Eins.
\end{equation}
Let $\Phi$ be a normalized eigenvector of $\widetilde{H}_a$
with minimal eigenvalue $\widetilde{E}_a$. Then
\begin{equation}\label{3.62}
\widetilde{E}_a = \langle \Phi | \widetilde{H}_a \Phi \rangle
=
\langle {\cal J}_a \Phi | {\cal B}_a(H_1) | {\cal J}_a \Phi \rangle
\ge \widetilde{\widetilde{E}}_a,
\end{equation}
where $\overline{E}_a$ is the minimal eigenvalue of
${\cal B}_a(H_1)$. Since the ground state energy of non--interacting
bosons is additive, we obtain further
\begin{equation}\label{3.63}
\overline{E}_a = a E_{\small min}(1)
=
a (\textstyle\frac{1}{4} jN + j_{\small min}  -j )
\end{equation}
and
\begin{equation}\label{3.64}
\widetilde{H}_a \ge \widetilde{E}_a \; \Eins \ge
a (\textstyle\frac{1}{4} jN + j_{\small min}  -j )\; \Eins.
\end{equation}
Using (\ref{3.61}) the final result is
\begin{equation}\label{3.65}
H_a \ge \left( \textstyle\frac{1}{4} Nj - a(j-j_{\small min})\right) \; \Eins.
\end{equation}


\section{Discussion}
In the above proof the two parts of the Hamiltonian acoording to
$H=H^Z+H^{XY}$ are considered separately. Thus this part of the proof could be
immediately generalized to the $XXZ$-model given by
\begin{equation}\label{3.66}
H(\Delta)=\Delta H^Z+H^{XY},\quad \Delta>0,
\end{equation}
similarly as in \cite{SSRS}. However, the considerations in section 2 concerning
the homogeneous gauge and in section 3.1 concerning the reduction to the
case $s=\frac{1}{2}$ presuppose an isotropic Hamiltonian. Hence an immediate
generalisation to the $XXZ$-model on the basis of the above proof
is only possible for weakly homogeneous systems and $s=\frac{1}{2}$. Compared
with the result in \cite{SSRS} this means that the condition
$J_{\mu\nu}\in\{0,J\},\quad J>0$ appearing in \cite{SSRS}
can be weakened to $J_{\mu\nu}\ge 0$. \\

As already pointed out, the above inequality (\ref{3.65}) is intended to apply for small values
of $a$, i.~e.~large values of $M=Ns-a$. For small $M$ much better estimates are
known \cite{SSL1}. However, for small $a$ the inequality cannot be improved
since there are examples where equality holds in (\ref{3.65}) for a couple of values of $a$,
e.~g.~$a=0,\ldots,\frac{N}{9}$, see \cite{SSRS} and \cite{SHSRS}.\\

Notwithstanding the construction of independent magnon states in particular examples,
the proof of (\ref{3.19}) anew establishes that the
Heisenberg Hamiltonian is only equivalent to the Hamiltonian of a Bose gas of magnons
if additional interaction terms are considered, see also \cite{BB}.
Apart from the repulsion term (\ref{3.55}) in the case of
AF-coupling an infinite repulsion term would have to be introduced which guarantees that no site is
occupied by more than one magnon (in the case $s=\frac{1}{2}$). Thus magnons appear as bosons
additionally satisfying the Pauli exclusion principle. The reader may ask why magnons are not rather
considered as fermions, for which the exclusion principle is automatically satisfied. The reason not
to do this is that the interchange of fermions at different sites would sometimes
produce factors of $-1$ which cannot be controlled, at least generally. For special topologies,
e.~g.~spin rings or chains the independent fermion concept works well and yields the
exact solution of the spin $\frac{1}{2}$ $XY$-model, see \cite{LSM}.

\section*{Acknowledgement}
The author is indepted to A.~Honecker for carefully reading a draft of the present
manuscript and giving valuable hints.


\section*{References}

\begin{thebibliography}{99}


\bibitem{Mat} D.~C.~Mattis, {\sl The Theory of Magnetism I}, Springer, Berlin,
	Heidelberg, New York (1981)
\bibitem{JM} J.~Schnack,  M.~Luban, Rotational modes in molecular magnets with
	antiferromagnetic Heisenberg exchange, 	Phys.~Rev.~{\bf B63}, 014418 (2001)
\bibitem{SSL1} H.~-J.~Schmidt, J.~Schnack,  M.~Luban, Bounding and Approximating
	parabolas for the spectrum of Heisenberg spin systems, Europhys. Lett.
	{\bf 55} (1), p. 105-111 (2001)
\bibitem{Wal} O.~Waldmann, Comment on ``Bounding and Approximating
	parabolas for the spectrum of Heisenberg spin systems" by
	H.~-J.~Schmidt, J.~Schnack and  M.~Luban, Europhys. Lett.~{\bf 57} (4), p. 618-619 (2002)
\bibitem{SSL2} H.~-J.~Schmidt, J.~Schnack,  M.~Luban, Reply to the
	Comment by O.~Waldmann on ``Bounding and Approximating
	parabolas for the spectrum of Heisenberg spin systems", Europhys. Lett.
	{\bf 57} (4), p. 620-621 (2002)
\bibitem{SSRS}J.~Schnack, H.~-J.~Schmidt, J.~Richter, J.~Schulenburg,
	Independent magnon states on magnetic polytopes,
          Eur.~Phys.~J.~{\bf B 24}. p. 475-481 (2001)
\bibitem{SHSRS}J.~Schulenburg, A.~Honecker, J.~Schnack, J.~ Richter, H.~-J.~Schmidt,
	Macroscopic magnetization jumps due to independent
          magnons in frustrated quantum spin lattices,
          Phys.~Rev.~Lett. (2002) accepted
\bibitem{Mil} F.~Mila, Ladders in a magnetic field: a strong coupling approach,
	Eur.~Phys.~J.~{\bf B6}, p.201-205 (1998)
\bibitem{MHSKU} E.~M\"uller--Hartmann, R.~R.~P.~Singh, Ch.~Knetter, G.~S.~Uhrig,
        Exact Demonstration of Magnetization Plateaus and First Order Dimer-Néel
        Phase Transitions in a Modified Shastry-Sutherland Model for $SrCu_2(BO_3)_2$,
        Phys.~Rev.~Lett.~{\bf 84}, p. 1808-1811 (2000)
\bibitem{HMT} A.~Honecker, F.~Mila, M.~Troyer, Magnetization plateaus and jumps
	in a class of frustrated ladders: A simple route to a complex behaviour,
        Eur.~Phys.~J.~{\bf B15}, p.227-233 (2000)
\bibitem{KOK} A.~Koga, K.~Okunishi, N.~Kawakami, First-order quantum phase transition
	in the orthogonal-dimer spin chain, Phys.~Rev.~{\bf B62}, p. 5556-58 (2000)
\bibitem{SR} J.~Schulenburg, J.~Richter, Infinite series of magnetization plateaus
        in the frustrated dimer-plaquette chain, Phys.~Rev.~{\bf  B 65} (5), 054420 (2002)
\bibitem{CB} E.~Chattopadhyay, I.~Bose, Magnetization of coupled spin clusters
	in ladder geometry, cond.mat./0107393
\bibitem{BB} E.~G.~Batyev, L.~S.~Braginski\u{\i}, Antiferromagnet in a Strong Magnetic Field:
	Analogy with Bose Gas, Soviet Physics JETP {\bf 60}, p. 781-786 (1984)
\bibitem{Glu} S.~Gluzman, Two-Dimensional Quantum Antiferromagnet in a Strong Magnetic Field,
	Z.~Phys.~{\bf B90}, p. 313-318, (1993)
\bibitem{SSS} S.~Sachdev, T.~Senthil, R.~Shankar, Finite-Temperature Properties of Quantum
	Antiferromagnets in a Uniform Magnetic Field in One and Two Dimensions, Phys.~Rev.~{\bf B50},
	p. 258-272 (1994)
\bibitem{LSM} E.~Lieb, T.~Schulz, D.~Mattis, Two Soluble Models of an Antiferromagnetic Chain,
	Ann.~Phys.~{\bf 16}, No.3, p.407-466 (1961)


\end{thebibliography}
\end{document}